\documentclass[a4paper,11pt]{article}
\usepackage{pos}
\usepackage{graphicx}
\usepackage{subcaption}
\usepackage[left]{lineno}
\title{Measurement of the Iron Spectrum with the MAGIC Telescopes}
\author*[a]{M. Molero}
\author[b]{S. Mangano, C. Delgado}
\author[]{for the MAGIC Collaboration}

\affiliation[a]{Instituto de Astrof\'{i}sica de Canarias (IAC) and Dpto. de Astrof\'{i}sica, Universidad de La Laguna, E-38200, La Laguna, Tenerife, Spain}

\affiliation[b]{Centro de Investigaciones Energéticas, Medioambientales y Tecnológicas (CIEMAT),
Avda. Complutense 40, Madrid, Spain}

\emailAdd{mmolerog@iac.es}
\abstract{Iron cosmic rays represent the most abundant heavy nuclei at energies above 1 TeV, with their production thought to be primarily originated by astrophysical sources. Therefore, measuring the iron spectrum provides crucial insights into the origin, acceleration, and propagation mechanisms of cosmic rays. While recent results from space-based detectors have revealed unexpected energy dependences in the GeV-TeV range, these measurements are limited by low statistics at higher energies. At energies above a few TeV, ground-based detectors, such as the Major Atmospheric Gamma Imaging Cherenkov (MAGIC) telescopes, become more effective due to their large collection areas, enabling them to extend and complement the capabilities of space-borne instruments. In this work, we apply the so-called direct Cherenkov technique, which accounts for the radiation emitted by charged particles before the cascade develops in the atmosphere, with MAGIC to identify iron-induced air showers and distinguish them from those produced by lighter cosmic-ray species. }

\ConferenceLogo{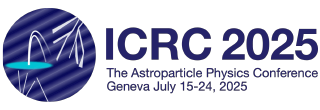}

\FullConference{39th International Cosmic Ray Conference (ICRC2025)\\
 15–24 July 2025\\
Geneva, Switzerland\\}
\begin{document}
\maketitle

\section{Introduction}

The all-particle cosmic-ray energy spectrum extends over more than ten orders of magnitude and exhibits several notable features. However, the exact sources and acceleration mechanisms of cosmic rays remain only partially understood. Precise measurements of their elemental composition are essential to constrain models of their origin and propagation. While direct detection experiments, such as balloon-borne or satellite missions, provide excellent charge resolution, they are limited in their energy range due to their relatively small collection areas. In contrast, ground-based air-shower experiments can access higher energies but generally suffer from poorer charge resolution.

A complementary technique for studying cosmic rays with Imaging Atmospheric Cherenkov Telescopes (IACTs) was first proposed in \cite{Ref1}. This method relies on the detection of Direct Cherenkov (DC) light, a short flash of Cherenkov radiation emitted by charged particles as they pass through the upper part of the atmosphere before initiating an extensive air shower (EAS). Since DC light is produced at higher altitudes than the EAS, it appears more focused in the IACT camera and can be identified as a bright pixel close to the shower direction. Furthermore, the intensity of DC light scales with the square of the particle’s charge ($Z^2$), making it particularly effective for identifying heavy elements such as iron, which dominates the cosmic-ray composition above a few TeV.

Previous measurements by H.E.S.S. \cite{Ref2} and VERITAS \cite{Ref3} have successfully applied this technique to reconstruct the iron flux at energies above 10 TeV.

In this work, we present the first measurement of DC light using the MAGIC telescopes. We describe the analysis method developed to isolate the DC signal and assess the systematic uncertainties in order to derive the energy spectrum.

\section{The MAGIC Telescopes}

MAGIC is a stereoscopic system consisting of two Imaging Atmospheric Cherenkov Telescopes (IACTs), located at the Roque de los Muchachos Observatory in La Palma, Spain (28.8°N, 17.8°W), at an altitude of approximately 2200 meters. The two telescopes, positioned 85 meters apart, each feature a 17-meter diameter reflector and a camera with a 3.5° field of view (FoV). MAGIC is primarily dedicated to perform gamma-ray studies, covering an energy range from around 20 GeV up to beyond 100 TeV. 

Although MAGIC is primarily designed for gamma-ray astronomy, the telescopes also detect a large number of cosmic-ray–induced air showers, which constitute the dominant background in gamma-ray observations. These cosmic rays, mostly protons and heavier nuclei, interact with the Earth's atmosphere to produce extensive air showers that with dedicated analysis techniques, can provide valuable information about the cosmic-ray spectrum at high energies. This also offers an opportunity for complementary astrophysical studies using the same instrumentation.

A detailed description  about trigger, data acquisition, and performance of the system can be found in \cite{Ref4}.

\section{Datasets and MC simulations}

The data used in this analysis were processed using the standard MAGIC analysis software (MARS) \cite{Ref5} to calibrate, perform the image cleaning and reconstruct the stereo Hillas parameters \cite{Ref6}. Special attention was given to preserving the intensity and arrival time information for each pixel surviving the image cleaning, as these parameters are crucial for this work but are not retained by default in the standard processing pipeline. In addition, to reduce potential systematic uncertainties, only data collected under good atmospheric conditions and with zenith angles less than 35$^\circ$ were selected. Table \ref{tab:data} summarizes the sources included in this analysis along with their respective exposure times after applying quality cuts. The total effective exposure time amounts to 305 hours.

The present work relies on Monte Carlo (MC) simulations to estimate the signal selection efficiency, energy reconstruction and collection areas, as well as the background rejection. In total, six different elements were produced using the CORSIKA 6.990 package \cite{Ref7} with the QGSJET-II-04 \cite{Ref8} hadronic interaction model for high energies and UrQMD 1.3.1 \cite{Ref9} for low energies. Iron was simulated to represent the signal component. Background simulations were grouped into charge (Z) bands: light-Z elements (proton and helium) and mid-Z elements (oxygen, magnesium, and silicon). A summary of the simulated elements and the number of generated events is presented in Table~\ref{tab:mc}.

\begin{table*}[h]
\centering
\begin{minipage}{0.48\textwidth}
\centering
\caption{Exposure times for each source used in this analysis}
\begin{tabular}{lc}
\hline
\hline
\textbf{Source} & \textbf{Exposure Time [h]} \\
\hline
M15         & 54.5 \\
M87         & 60.3 \\
Draco       & 34.1 \\
Cygnus-X3   & 36.7 \\
Crab Nebula & 72.3 \\
Mrk421      & 47.1 \\
\hline
\end{tabular}
\label{tab:data}
\end{minipage}
\hfill
\begin{minipage}{0.48\textwidth}
\centering
\caption{Monte Carlo productions for various nuclei. Particle type, charge (Z), and number of generated events (Gen. Events) are shown.}
\begin{tabular}{lccc}
\hline
\hline
\textbf{Particle} & \textbf{Z} & \textbf{Gen. Events} [$\times 10^{6}$] \\
\hline
Proton     & 1  & 1.5 \\
Helium     & 2  & 1.2 \\
Oxygen     & 8  & 0.7 \\
Magnesium  & 12 & 0.75 \\
Silicon    & 14 & 2.0 \\
Iron       & 26 & 10.0 \\
\hline
\end{tabular}
\label{tab:mc}
\end{minipage}
\end{table*}

\section{Analysis Method}

\subsection{Energy Reconstruction}

The energy reconstruction process relies on the novel technique developed for the MAGIC telescopes based on the machine learning algorithm Random Forest (RF) \cite{Ref10}. The method uses as inputs image parameters from the standard Hillas parametrization as well as optimized parameters for the DC-light detection. Iron MC is used to train and test the RF, providing a less than 10$\%$ bias and less than 15$\%$ resolution for DC-light events from 10 to 100 TeV. The outputs of the training are used in real data and background MCs.

\subsection{Data Selection}

Iron-induced showers producing DC light are identified by means of a two-step cut-based selection procedure: quality cuts to ensure reliable event reconstruction, and identification cuts to isolate the DC signal.

The first part, ensures a well-reconstructed stereo event in both MAGIC telescopes. These include minimum values on image parameters such as size, width, and length, derived from stereo Hillas reconstruction. In addition, to reduce edge effects and avoid truncated images, events with a center of gravity (c.o.g.) near the camera edge are excluded.

Identification cuts aim to detect DC light as a distinct, bright pixel located near the reconstructed shower direction. Two estimators are used, following the approach developed by H.E.S.S. \cite{Ref2}. The first one, DC-ratio ($Q_{DC}$), is defined as:

\begin{equation}
    Q_{DC} = \frac{<I_{ngh}>}{I_{max}}
\end{equation}
where $I_{max}$ is the intensity of the brightest pixel, and $<I_{ngh}>$ is the average intensity of the neighboring pixels. This ratio captures the contrast between a potential DC signal and its surroundings.

\begin{figure}[htbp]
    \centering
    \begin{subcaptionbox}{\label{fig:qdca}}[0.48\linewidth]
        {\includegraphics[width=\linewidth]{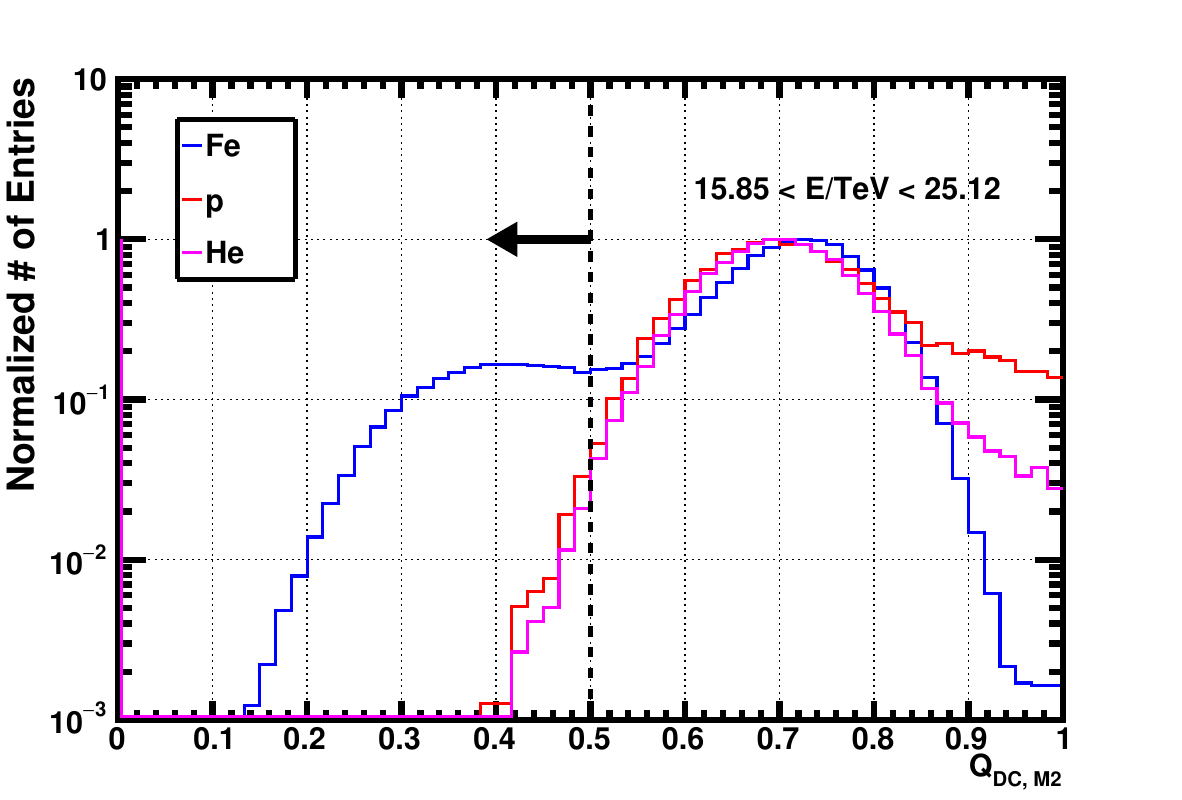}}
    \end{subcaptionbox}
    \hfill
    \begin{subcaptionbox}{\label{fig:qdcb}}[0.48\linewidth]
        {\includegraphics[width=\linewidth]{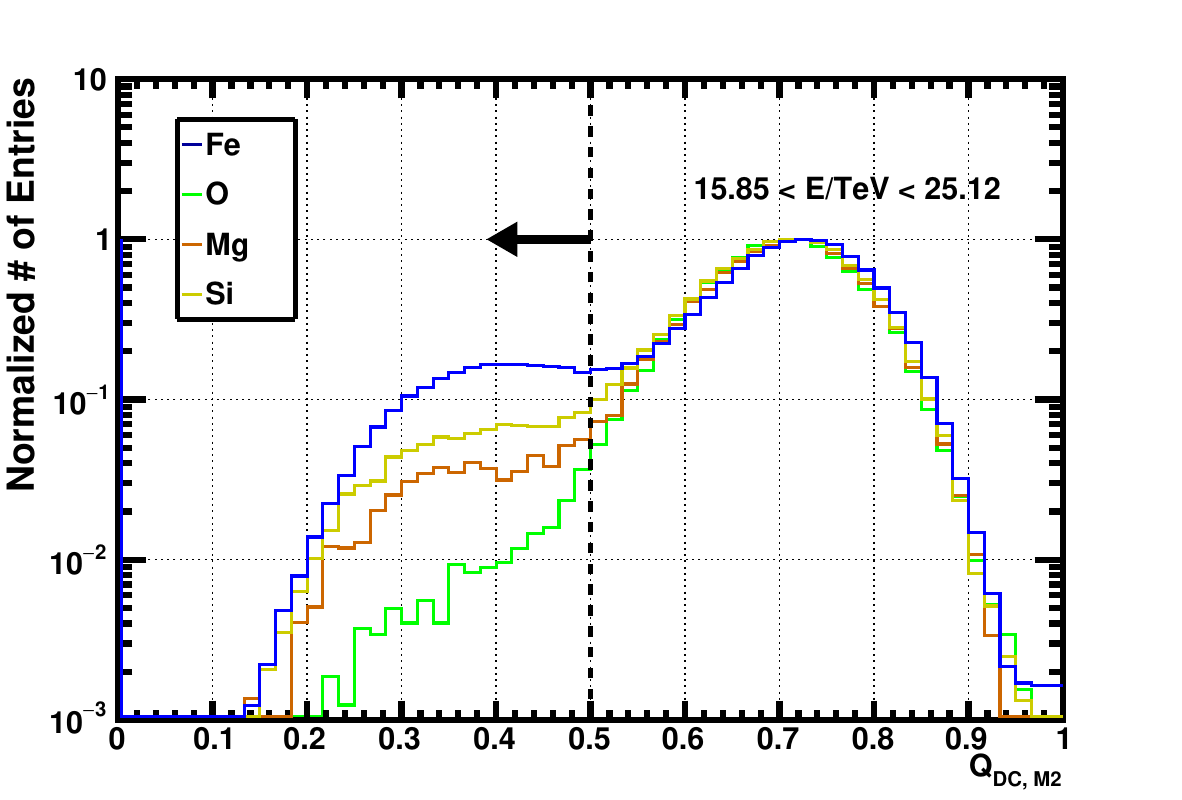}}
    \end{subcaptionbox}
    \caption{Direct Cherenkov Ratio ($Q_{DC}$) for M2 and for the energy range 15.85 < E/TeV < 25.12. Figure (a) shows the comparisons of $Q_{DC}$ for iron simulated events with respect to p and he simulations. Figure (b) compares iron with oxygen, magnesium and silicon. }
    \label{fig:qdc}
\end{figure}

As an example, $Q_{DC}$ in M2 and for the energy range 15.85 < E/TeV < 25.12 after applying quality cuts is shown in Figure \ref{fig:qdc} for the signal and different background MCs. Figure \ref{fig:qdca} shows the comparison of $Q_{DC}$ for iron with respect to the light-Z background component (proton and helium), while Figure \ref{fig:qdcb} presents iron versus the mid-Z component (oxygen, magnesium and silicon). As can be seen, this estimator is able to efficiently reject proton and helium events, but it is less effective for mid-Z background. A cut of $Q_{DC} < 0.5$ is applied to keep candidate iron events.

The second estimator is the DC-light intensity ($I_{DC}$) defined as:

\begin{equation}
    I_{DC} = I_{max} - <I_{ngh}>
\end{equation}
where $I_{max}$ and $<I_{ngh}>$ are computed as previously explained. 

\begin{figure}[htbp]
    \centering
    \begin{subcaptionbox}{\label{fig:idca}}[0.48\linewidth]
        {\includegraphics[width=\linewidth]{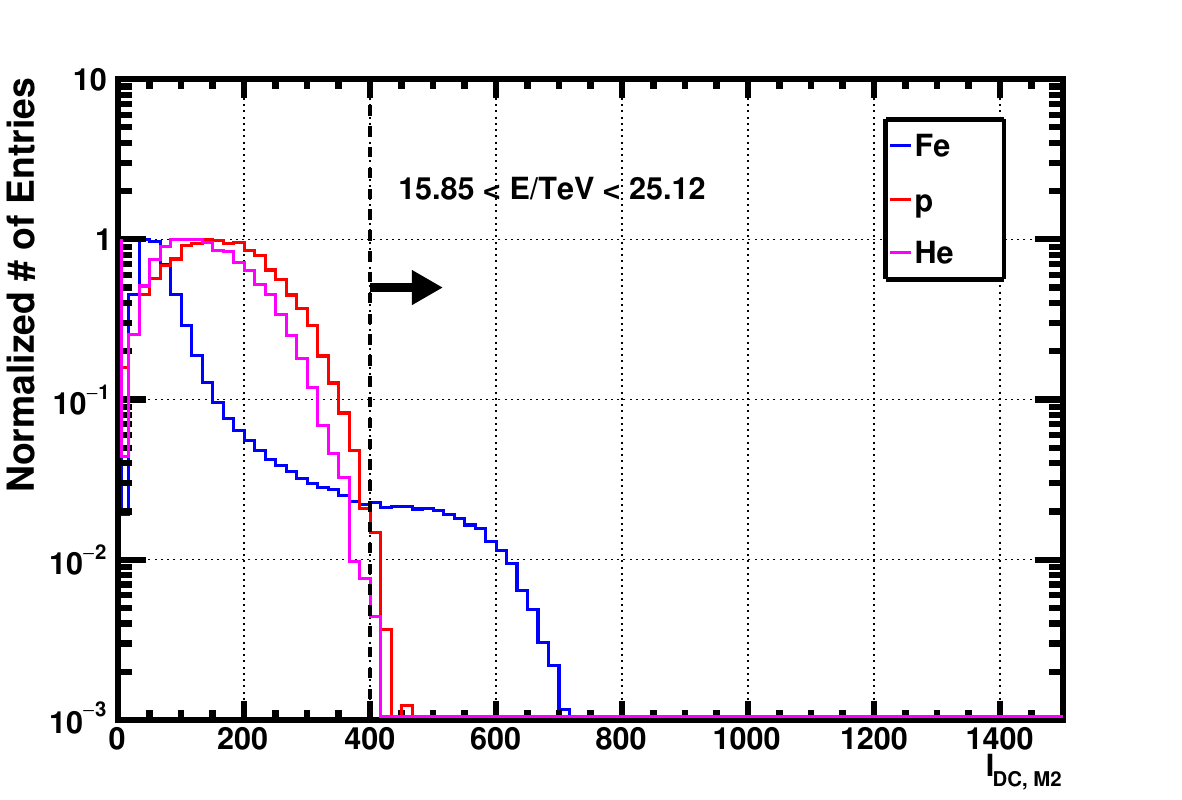}}
    \end{subcaptionbox}
    \hfill
    \begin{subcaptionbox}{\label{fig:idcb}}[0.48\linewidth]
        {\includegraphics[width=\linewidth]{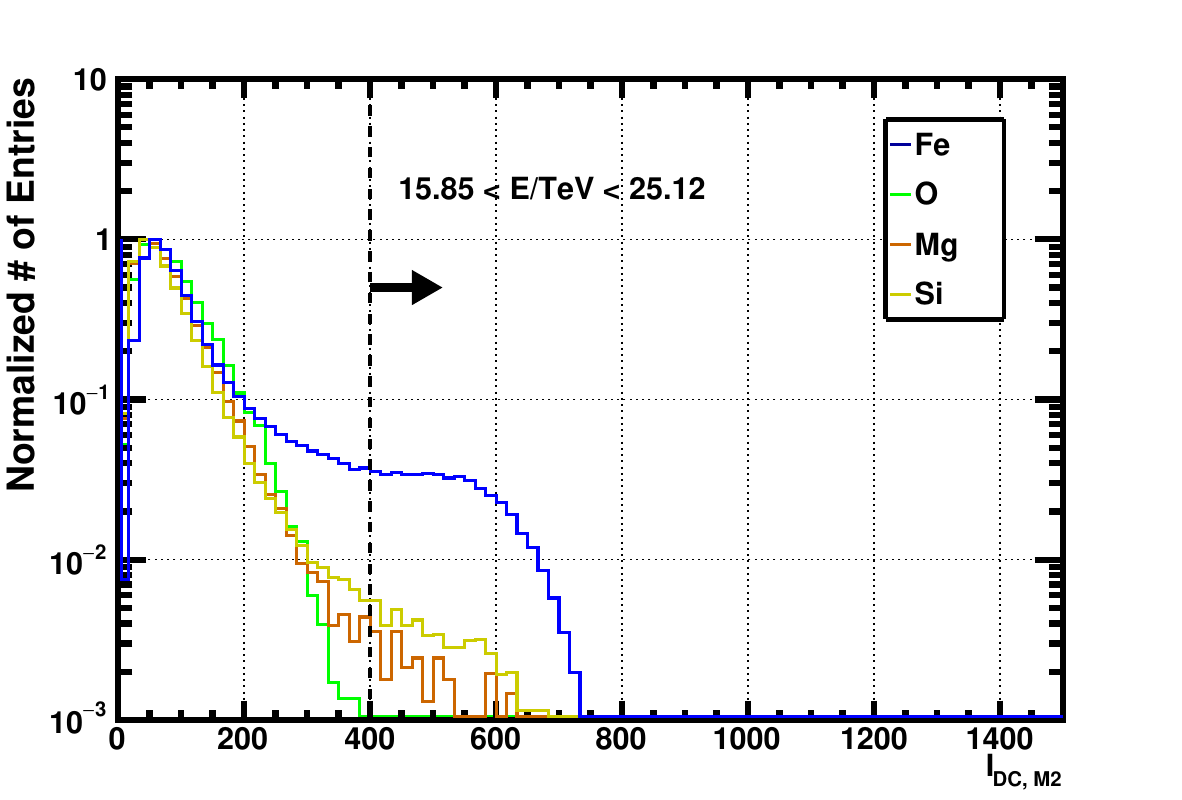}}
    \end{subcaptionbox}
    \caption{Intensity of the Direct Cherenkov Light ($I_{DC}$) for M2 and for the energy range 15.85 < E/TeV < 25.12. Figure (a) shows the comparisons of $I_{DC}$ for iron simulated events with respect to p and he simulations. Figure (b) compares iron with oxygen, magnesium and silicon. }
    \label{fig:idc}
\end{figure}

An example of this estimator in M2 and for the same energy range as before after applying quality cuts is shown in Figure \ref{fig:idc} for the signal and background MCs. Figure \ref{fig:idca} and \ref{fig:idcb} show the comparison of iron with proton and helium, and with oxygen, magnesium and silicon, respectively. A cut of $I_{DC} > 400$ is applied to improve separation from mid-Z nuclei.

Finally, to avoid detector saturation effects, the intensity of the brightest pixel is clipped to 900 photoelectrons (p.e.) in both data and MC after the identification.

\begin{figure}[htbp]
    \centering
    \begin{subcaptionbox}{\label{fig:fig1}}[0.48\linewidth]
        {\includegraphics[width=\linewidth]{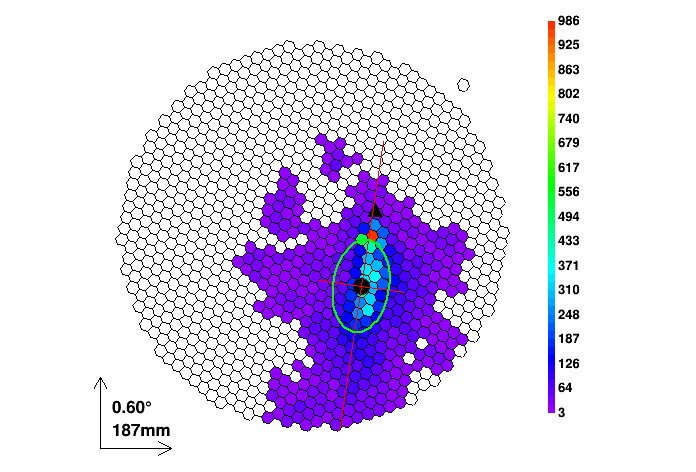}}
    \end{subcaptionbox}
    \hfill
    \begin{subcaptionbox}{\label{fig:fig2}}[0.48\linewidth]
        {\includegraphics[width=\linewidth]{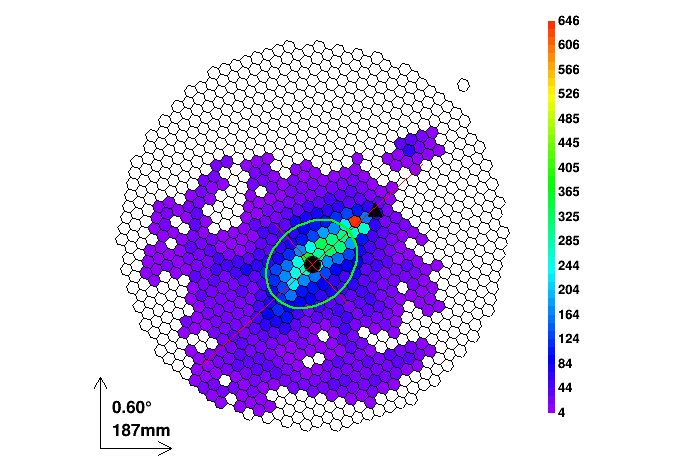}}
    \end{subcaptionbox}
    \caption{Measured iron event shower in real data with evident Direct Cherenkov light in M1 (a) and M2 (b) . The black triangle and dot show the reconstructed shower direction and c.o.g. respectively, whereas the two red perpendicular lines within the ellipse correspond to the Width and Length obtained from the standard Hillas parameterization. The reconstructed energy of this event is 30.30 TeV. }
    \label{fig:evt}
\end{figure}

The combined application of the identification cuts in $Q_{DC}$ and $I_{DC}$ in both telescopes provide a negligible contribution of light-Z nuclei, as well as a reduced mid-Z contribution. An example event from real data, identified as an iron candidate, is shown in Figure \ref{fig:evt}. In total, 287 DC-light events are identified in the energy range from 10 to 60 TeV.

%The bin-to-bin migrations due to the limited energy resolution of the detector are taken into account via a regularized likelihood minimization method using a reference and the migration matrix.

\subsection{Systematic Uncertainties}

A preliminary study of the systematic uncertainties affecting the measurement has been carried out. The following sources have been considered:

\begin{itemize}
    \item Typical MAGIC systematic effects and energy reconstruction: 
    \(\sigma_{\text{ene}} = 15\%\)
    
    \item Hadronic interaction model uncertainty in CORSIKA \cite{Ref11}: 
    \(\sigma_{\text{had}} = 10\%\)
    
    \item Influence of nuclei with \(Z \leq 14\): 
    \(\sigma_{Z<14} \leq 7\%\) 
    
    \item Influence of nuclei with \(Z > 14\): 
    \(\sigma_{Z>14} \leq 25\%\) 
\end{itemize}

The total systematic uncertainty is calculated by adding the individual contributions in quadrature, resulting in $\sigma_{tot}=32\%$.

\section{Conclusions}

We have presented the first detection of Direct Cherenkov (DC) light using the MAGIC telescopes, and described the methodology required to derive the cosmic-ray iron spectrum. In particular, a dedicated event selection method was followed, incorporating optimized quality and identification cuts to isolate iron-induced showers with DC-light signatures while suppressing light and mid-Z background nuclei. A total of 287 DC-light events were identified in the real data. Systematic uncertainties were preliminarily assessed, with the total contribution estimated at 32$\%$. These results confirm the feasibility of DC-light measurements with MAGIC and open the door to future cosmic-ray composition studies. Continued efforts, including the use of larger datasets and more advanced analysis techniques, will further improve the precision and extend the accessible energy range of this method.

\section{Acknowledgements}
We would like to thank the Instituto de Astrof\'{\i}sica de Canarias for the excellent working conditions at the Observatorio del Roque de los Muchachos in La Palma. The financial support of the German BMBF, MPG and HGF; the Italian INFN and INAF; the Swiss National Fund SNF; the grants PID2019-107988GB-C22, PID2022-136828NB-C41, PID2022-137810NB-C22, PID2022-138172NB-C41, PID2022-138172NB-C42, PID2022-138172NB-C43, PID2022-139117NB-C41, PID2022-139117NB-C42, PID2022-139117NB-C43, PID2022-139117NB-C44, CNS2023-144504 funded by the Spanish MCIN/AEI/ 10.13039/501100011033 and "ERDF A way of making Europe; the Indian Department of Atomic Energy; the Japanese ICRR, the University of Tokyo, JSPS, and MEXT; the Bulgarian Ministry of Education and Science, National RI Roadmap Project DO1-400/18.12.2020 and the Academy of Finland grant nr. 320045 is gratefully acknowledged. This work was also been supported by Centros de Excelencia ``Severo Ochoa'' y Unidades ``Mar\'{\i}a de Maeztu'' program of the Spanish MCIN/AEI/ 10.13039/501100011033 (CEX2019-000920-S, CEX2019-000918-M, CEX2021-001131-S) and by the CERCA institution and grants 2021SGR00426 and 2021SGR00773 of the Generalitat de Catalunya; by the Croatian Science Foundation (HrZZ) Project IP-2022-10-4595 and the University of Rijeka Project uniri-prirod-18-48; by the Deutsche Forschungsgemeinschaft (SFB1491) and by the Lamarr-Institute for Machine Learning and Artificial Intelligence; by the Polish Ministry Of Education and Science grant No. 2021/WK/08; and by the Brazilian MCTIC, CNPq and FAPERJ.

\end{document}